\documentclass[preprintnumbers,amsmath,amssymb,showpacs]{revtex4}
\usepackage{epsfig}
\usepackage{color}
\usepackage[colorlinks,urlcolor=blue,citecolor=blue]{hyperref}
\begin{document}
\setlength{\voffset}{1.0cm}
\title{Semiclassical time crystal in the chiral Gross-Neveu model}
\author{Michael Thies\footnote{michael.thies@gravity.fau.de}}
\affiliation{Institut f\"ur  Theoretische Physik, Universit\"at Erlangen-N\"urnberg, D-91058, Erlangen, Germany}
\date{\today}
\begin{abstract}
In the limit of a large number of flavors, the ground state of the chiral Gross-Neveu model at finite fermion density exhibits a spatially periodic
mean field in the form of the chiral spiral, thereby breaking translational invariance. Here we show that the ground state of the same model at finite 
fermion current density gives rise to a mean field which is periodic in time, a temporal chiral spiral. Since the current density is the same as 
axial charge density in two dimensions and axial charge is conserved, this may serve as an example of a time crystal. More specifically, as mean field theory is invoked
in the large $N$ limit, we are dealing with a semiclassical time crystal. 
\end{abstract}
\maketitle
Recent speculations about time crystals in quantum mechanics \cite{L1,L2,L3,L4,L5,L6,L7,L8}, i.e., systems whose ground state displays a periodic time dependence,
have incited us to revisit the chiral Gross-Neveu model \cite{L9}. This is 
a relativistic quantum field theory in 1+1 dimensions, also referred to as two-dimensional Nambu--Jona-Lasinio (NJL$_2$) model \cite{L10}, with Lagrangian
\begin{equation}
{\cal L}_{\rm NJL} =  \sum_{i=1}^N \bar{\psi}^{(i)} i\partial \!\!\!/ \psi^{(i)} + \frac{g^2}{2} \left[\left( \sum_{i=1}^N \bar{\psi}^{(i)} \psi^{(i)} \right)^2 +
\left( \sum_{i=1}^N \bar{\psi}^{(i)} i \gamma_5\psi^{(i)} \right)^2\right] \, .
\label{A1}
\end{equation}
Its most characteristic property is the U(1)$\otimes$U(1) chiral symmetry,
\begin{equation}
\psi \to e^{i\alpha} \psi, \quad \psi \to e^{i\beta \gamma_5} \psi,
\label{A2}
\end{equation}
leading to conserved vector and axial vector currents
\begin{equation}
\partial_{\mu} j^{\mu} = 0, \quad \partial_{\mu} j_5^{\mu} = 0
\label{A3}
\end{equation} 
and corresponding conserved charges
\begin{equation}
\partial_t Q = \partial_t \int dx j^0 = 0, \quad \partial_t Q_5 = \partial_t \int dx j_5^0 = 0.
\label{A4}
\end{equation}
In 1+1 dimensions, the currents $j$ and $j_5$ are closely related,
\begin{equation}
j^0 = j_5^1 = \psi^{\dagger} \psi, \quad j^1 = j_5^0 = \psi^{\dagger} \gamma_5 \psi.
\label{A5}
\end{equation}
For a demonstration of the conservation laws (\ref{A3}) and a discussion of their implications, see Refs.~\cite{L10a,L10b}. 
Due to Eq.~(\ref{A4}) it is equally legitimate to study the system in a sector with a definite vector charge $Q$ (sum of right- and left-handed fermions) as in a sector with a 
definite axial charge $Q_5$ (difference of right- and left-handed fermions). In the large $N$ limit, mean field theory becomes exact and one can solve the NJL$_2$ model 
using the relativistic Hartree-Fock (HF) approach (for an introduction, see \cite{L11}). The ground state at finite fermion density has turned out to be a paradigm for spontaneous breakdown of
translational symmetry in space. At all densities, the mean field assumes the form of a helix in the space spanned by the scalar
potential $S=-Ng^2 \langle \bar{\psi}\psi\rangle$, the pseudoscalar potential $P=-Ng^2 \langle \bar{\psi} i \gamma_5 \psi \rangle$ and
the space coordinate $x$, the ``chiral spiral" \cite{L12}. The analogous question for a system with finite axial charge has not yet
been addressed in the literature, to the best of our knowledge. We will show that this actually yields a candidate for a time
crystal, with a chiral spiral winding in the time direction rather than the space direction. Notice that a temporal chiral spiral has
been discussed in the phenomenologically more relevant context of quantum chromodynamics (QCD) with external electromagnetic fields recently \cite{L13}.
The present example is significantly 
more elementary and has the advantage that we do not need any approximation other than the large $N$ limit, owing to the simplicity of the model. 
The strict large $N$ limit suppresses quantum fluctuations of the mean field, a prerequisite for escaping the Coleman-Mermin-Wagner theorem \cite{L13a,L13b}
which forbids breaking of a continuous symmetry in 1+1 dimensions.

In order to motivate our study, we find it instructive to start discussing finite density and finite axial density systems in parallel. As far as 
finite density is concerned, everything is of course well-known. It is nevertheless necessary to repeat several steps of the derivation here in
order to better understand the axial density case. We follow closely the original references \cite{L11,L12}.

Let us start with the naive approach where it is assumed that the mean field in the ground state is homogeneous in space and time. Without 
loss of generality, we can set $S=m, P=0$, where $m$ is the dynamical fermion mass. The HF equation
then reduces to the massive, free Dirac equation
\begin{equation}
\left( -i \gamma_5 \partial_x + m \gamma^0 \right) \psi = \epsilon \psi
\label{A6}
\end{equation}
and can trivially be solved. For later convenience, we write down the solutions in the representation where $\gamma^0 = -\sigma_1,
\gamma^1 = i \sigma_3, \gamma_5 =  -\sigma_2$,
\begin{eqnarray}
\psi_k^{(+)} & = & e^{ikx} u_k, \quad \epsilon = e_k := \sqrt{k^2+m^2},
\nonumber \\
\psi_k^{(-)} & = & e^{ikx} v_k, \quad \epsilon = - e_k ,
\label{A7}
\end{eqnarray}
with the normalized positive and negative energy spinors
\begin{equation}
u_k = \frac{1}{\sqrt{2} e_k} \left( \begin{array}{c} ik-m \\ e_k \end{array} \right), \quad 
v_k = \frac{1}{\sqrt{2} e_k} \left( \begin{array}{c} ik-m \\ -e_k \end{array} \right).
\label{A8}
\end{equation}
\begin{figure}[h]
\begin{center}
\epsfig{file=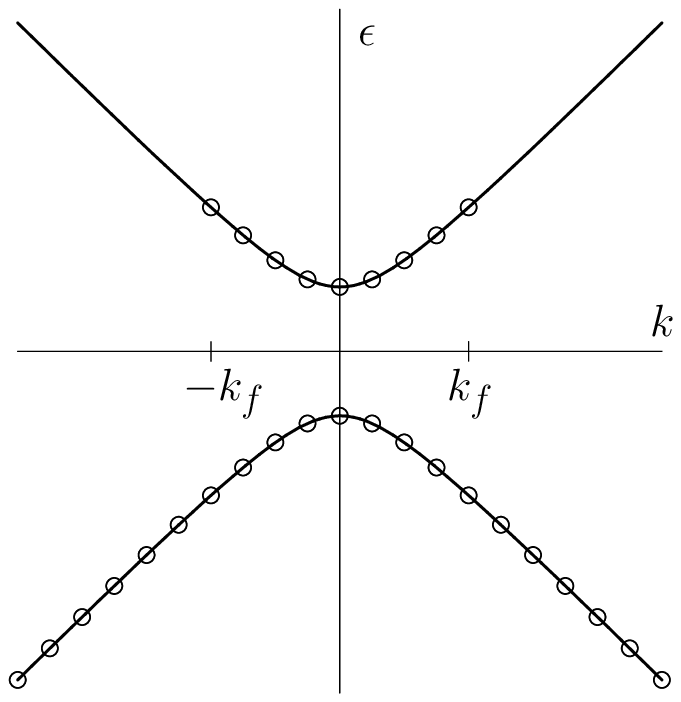,width=6cm,angle=0}
\caption{Filling of single particle states for HF calculation at finite density $\rho=k_f/\pi$ and homogeneous mean field.}
\label{Fig1}
\end{center}
\end{figure}
\begin{figure}[h]
\begin{center}
\epsfig{file=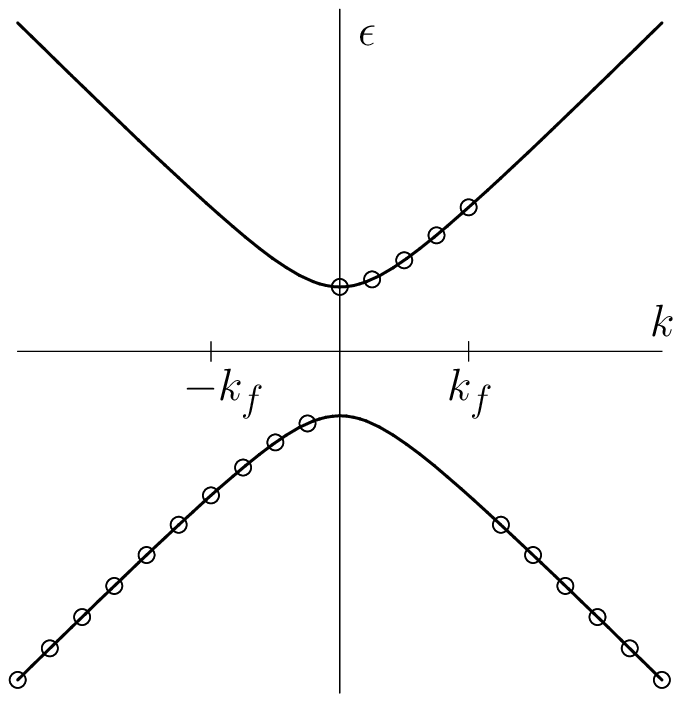,width=6cm,angle=0}
\caption{Filling of single particle states for HF calculation at finite axial density, Eq.~(\ref{A9}), and homogeneous mean field. Note that this configuration
has the same energy as the one shown in Fig.~\ref{Fig1}.} 
\label{Fig2}
\end{center}
\end{figure}
The crucial issue is how the single particle states are filled. In the finite density case, we occupy all negative energy states 
(the Dirac sea) as well as the positive energy states in the interval $[-k_f,k_f]$, the 1-dimensional Fermi sphere (Fig.~\ref{Fig1}). How do 
we fill the states in the case of a finite axial density? Starting from the vacuum with all negative energy 
states filled, we occupy in addition the positive energy states in the interval $[0,k_f]$ and empty the negative energy states in 
the same interval (Fig.~\ref{Fig2}). This yields equal densities of right moving fermions and left moving antifermions at minimal
cost of energy. 
Let us compute the density and axial density for these two cases (relative to the vacuum) 
\begin{eqnarray}
\frac{\rho}{N} & = & \int_{-k_f}^{k_f} \frac{dk}{2\pi} = \frac{k_f}{\pi},
\nonumber \\
\frac{\rho_5}{N} & = & 2 \int_0^{k_f} \frac{dk}{2\pi} \frac{k}{e_k} = \frac{\sqrt{k_f^2+m^2}-m}{\pi}.
\label{A9}
\end{eqnarray}
The HF ground state energy density is identical in both cases, namely
\begin{equation}
\frac{\cal E}{N} = - 2 \int_{k_f}^{\Lambda/2} \frac{dk}{2\pi} \sqrt{m^2+k^2} + \frac{m^2}{2Ng^2},
\label{A10}
\end{equation}
so that the following steps apply as well to finite density as to finite current. 
Renormalizing at $k_f=0$ (vacuum fermion mass $m_0$), we find
\begin{equation}
\frac{\cal E}{N} = - \frac{m^2}{4\pi} + \frac{k_f}{2\pi} \sqrt{k_f^2+m^2} + \frac{m^2}{2\pi}\ln \left( \frac{k_f + \sqrt{m^2+k_f^2}}{m_0} \right).
\label{A11}
\end{equation}
To determine the dynamical fermion mass at finite $k_f$, we minimize ${\cal E}$ with respect to $m$,
\begin{equation}
m \ln \left( \frac{k_f + \sqrt{m^2+k_f^2}}{m_0} \right) = 0,
\label{A12}
\end{equation}
with the solutions
\begin{equation}
m=0, \quad m=m_0 \sqrt{1-\frac{2 k_f}{m_0}} \quad \left( k_f < \frac{m_0}{2} \right).
\label{A13}
\end{equation}
The corresponding energy densities are 
\begin{eqnarray}
\left. \frac{\cal E}{N} \right|_{m=0} & = & \frac{k_f^2}{2\pi},
\nonumber \\
\left. \frac{\cal E}{N} \right|_{m\neq 0} & = &  - \frac{m_0^2}{4\pi} + \frac{k_f m_0}{\pi} - \frac{k_f^2}{2\pi} \quad \left(k_f < \frac{m_0}{2} \right).
\label{A14}
\end{eqnarray}
\begin{figure}[h]
\begin{center}
\epsfig{file=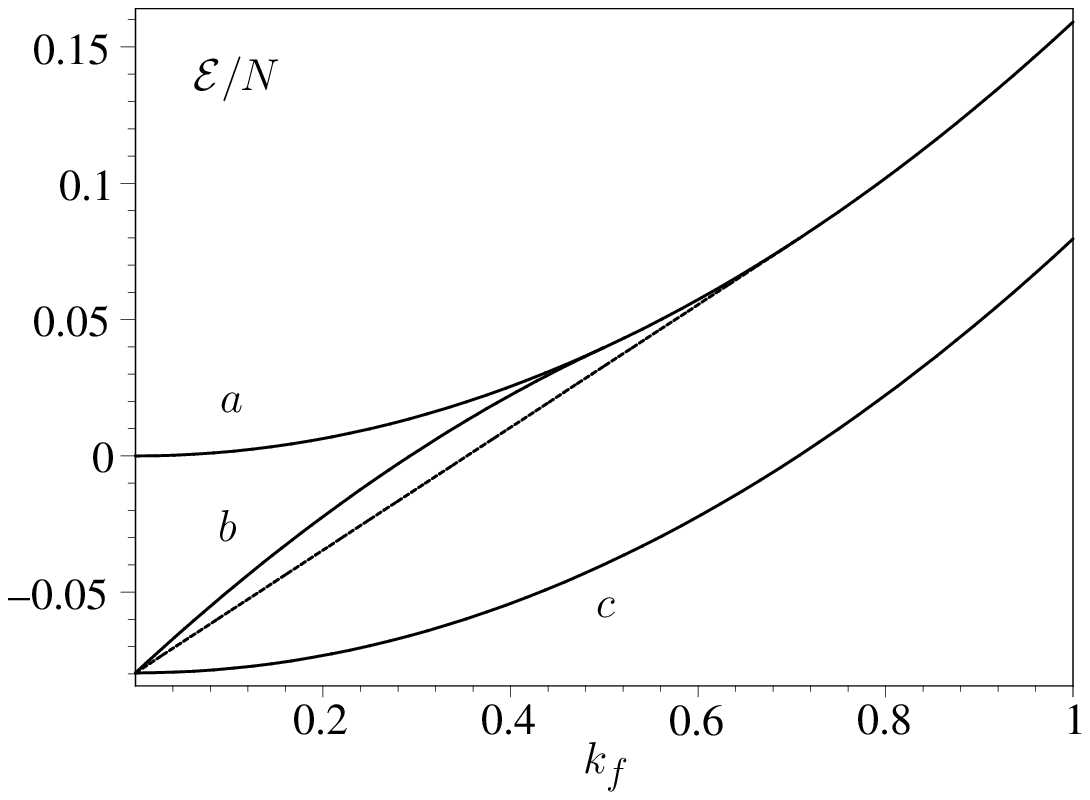,width=8cm,angle=0}
\caption{Energy density vs. $k_f$ for NJL$_2$ model at finite density or axial density, in units where $m_0=1$. Curve $a$: Chirally unbroken phase, 1st line of Eq.~(\ref{A14}). 
Curve $b$: Chirally broken phase with homogeneous HF potential, 2nd line of Eq.~(\ref{A14}). Curve $c$: spatial chiral spiral, Eq.~(\ref{A25}). The straight line is the mixed phase, 
Eq.~(\ref{A15}).}
\label{Fig3}
\end{center}
\end{figure}
As shown in Fig.~\ref{Fig3}, the curve ${\cal E}(k_f)$ with broken chiral symmetry ($m\neq 0$) has lower energy, but fails to be convex. This signals 
some instability of the homogeneous mean field. 
In the case of finite density, it is then energetically advantageous to restore chiral symmetry in part of the space and put massless fermions
in there, whereas the rest of space contains the chirally broken vacuum. The result of a variational calculation along these lines 
is a linear $k_f$ dependence 
\begin{equation}
\frac{\cal E}{N} = - \frac{m_0^2}{2\pi} + \frac{k_f m_0}{\sqrt{2}\pi} \quad \left( k_f < \frac{m_0}{\sqrt{2}}\right).
\label{A15}
\end{equation}
This leads to the scenario with a first order phase transition at $k_f= m_0/\sqrt{2}$. A study of the grand canonical potential
with chemical potential confirms this result and identifies the region $0<k_f<m_0/\sqrt{2}$ as mixed phase \cite{L14}. In the case of the current density,
it is difficult for us to envisage a mixed phase, since droplets of current are hard to reconcile with current conservation.
Since the HF solutions with homogeneous mean fields are anyway unstable, it does not seem worthwhile to try and solve this puzzle.

We now come to the true ground states of our systems, using again the HF approach. We first remind the reader about the finite density case and then turn to
finite axial density. The arguments leading to the chiral spiral can be phrased as follows. Write down the HF equation with a specific,
inhomogeneous ansatz for the mean field,
\begin{equation}
\left( -i \gamma_5 \partial_x + \gamma^0 m e^{-2iax\gamma_5} \right) \psi = \epsilon \psi.
\label{A16}
\end{equation} 
This chiral spiral mean field thus consists of oscillating scalar and pseudoscalar terms
\begin{equation}
\gamma^0 m e^{-2iax\gamma_5} = \gamma^0 S(x) + i \gamma^1 P(x), \quad S(x)= m \cos (2ax), \quad P(x) = -m \sin (2ax).
\label{A17}
\end{equation}
The two parameters $m,a$ determine the radius and the pitch of the chiral spiral. They have to be fixed through the self-consistency
condition and the value of the density. We solve Eq.~(\ref{A16}) by first performing a local chiral transformation
\begin{equation}
\psi = e^{iax\gamma_5} \tilde{\psi}.
\label{A18}
\end{equation}
This maps the HF equation onto the free, massive Dirac equation except for an overall shift in energy,
\begin{equation}
\left( -i \gamma_5 \partial_x + \gamma^0 m + a  \right) \tilde{\psi} = \epsilon \tilde{\psi}.
\label{A19}
\end{equation} 
Hence the solutions of the HF equation (\ref{A16}) are 
\begin{eqnarray}
\psi_k^{(+)} & = & e^{iax\gamma_5} e^{ikx} u_k, \quad \epsilon= a + e_k ,
\nonumber \\
\psi_k^{(-)} & = & e^{iax\gamma_5} e^{ikx} v_k, \quad \epsilon= a - e_k .
\label{A20}
\end{eqnarray}
We fill all $\psi_k^{(-)}$ states in the interval $k \in [-\Lambda'/2,\Lambda'/2]$ where $\Lambda'= \Lambda+2a$ and proceed to prove
self-consistency and determine the free parameters $m,a$. Subsequently we shall evaluate the fermion density and energy density of the HF state.
The scalar and pseudoscalar densities for the single particle state with quasi-momentum $k$,
\begin{eqnarray}
\bar{\psi}_k^{(-)}\psi_k^{(-)} & = & - \frac{m}{e_k} \cos (2ax),
\nonumber \\
\bar{\psi}_k^{(-)} i \gamma_5 \psi_k^{(-)} & = & \frac{m}{e_k} \sin(2ax),
\label{A21}
\end{eqnarray}
yield the following mean fields,
\begin{eqnarray}
S & = &  - Ng^2 \int_{-\Lambda'/2}^{\Lambda'/2} \bar{\psi}_k^{(-)} \psi_k^{(-)}
 =   \frac{Ng^2}{2\pi} \ln \left( \frac{\Lambda^2}{m^2} \right) m \cos (2ax),
\nonumber \\
P & = &  - Ng^2 \int_{-\Lambda'/2}^{\Lambda'/2} \bar{\psi}_k^{(-)}i \gamma_5 \psi_k^{(-)}
 =  - \frac{Ng^2}{2\pi} \ln \left( \frac{\Lambda^2}{m^2} \right) m \sin(2ax).
\label{A22}
\end{eqnarray}
Owing to the vacuum gap equation
\begin{equation}
\frac{Ng^2}{2\pi} \ln \frac{\Lambda^2}{m_0^2} = 1,
\label{A23}
\end{equation}
$S$ and $P$ agree with the input (\ref{A17}) for the choice $m=m_0$, the vacuum fermion mass, which we assume from now on. 
This proves that the chiral spiral solution is self-consistent. For the fermion density we find
\begin{equation}
\frac{\rho}{N} = \int_{-\Lambda'/2}^{\Lambda'/2} \frac{dk}{2\pi} = \frac{\rho_0}{N} + \frac{a}{\pi}
\label{A24}
\end{equation}
where $\rho_0$ is the divergent fermion density of the vacuum.
The energy density becomes
\begin{equation}
\frac{\cal E}{N} =  \int_{\Lambda'/2}^{\Lambda'/2}\frac{dk}{2\pi} \left(a -  \sqrt{k^2+m_0^2} \right) + \frac{m_0^2}{2Ng^2} = \frac{{\cal E}_0}{N} + \frac{a^2}{2\pi}
\label{A25}
\end{equation}
where ${\cal E}_0$ is the vacuum energy density. Subtracting the vacuum contributions, the fermion density and energy density per flavor
of the chiral spiral are $a/\pi$ and $a^2/2\pi$, respectively. In order to compare this with the homogeneous HF solution,
we have to choose $a=k_f$ and plot the energy curve into Fig.~\ref{Fig3} with the well-known result. The chiral spiral yields the lowest energy states
at all densities, and chiral symmetry does not get restored anywhere. 

We now turn to the system with finite axial density or current density. The fact that the energy is not a convex curve as a function of axial density points to a similar
stability problem as in the finite density case, but the remedy must be different. The standard chiral spiral induces fermion density, but not current density.
The key feature which makes the chiral spiral work is the axial anomaly \cite{L15}. For our purpose, it may be phrased as follows: A local chiral rotation
 of the fermions in the vacuum by $e^{i \chi(x) \gamma_5}$ induces the fermion density
\begin{equation}
\rho(x) = \frac{\partial_x \chi(x)}{\pi}.
\label{A26}
\end{equation}
By covariance, a local chiral rotation with the ($x,t$)-dependent angle $\chi$ should induce 
a vector current and an axial vector current of the form
\begin{equation}
j^{\mu} = - \frac{\epsilon^{\mu}_{\ \nu}\partial^{\nu} \chi}{\pi}, \quad j_5^{\mu} = - \frac{\partial^{\mu} \chi}{\pi},
\label{A27}
\end{equation}
where we have used (\ref{A5}). Incidentally, the conservation law for the induced axial current,
\begin{equation}
\partial_{\mu} j_5^{\mu} = - \frac{\partial_{\mu}\partial^{\mu} \chi}{\pi} = 0,
\label{A28}
\end{equation} 
can be interpreted as Klein-Gordon equation for a massless Goldstone boson field, the would-be pion.
Eq.~(\ref{A27}) suggests that a linearly time dependent chiral spiral is what it takes to induce a constant axial charge density.
Notice that there is no contradiction between conservation of axial charge and the presence of the axial anomaly in the NJL$_2$
model. In 1+1 dimensional gauge theories like the Schwinger model \cite{L15a}, it is impossible to reconcile conservation of
axial charge and gauge invariance, as explained in detail in Refs.~\cite{L15b,L15c}. In our case, there is no such conflict because
we do not gauge any of the global symmetry transformations (\ref{A2}).
 
It is well known how to generalize the HF formalism to time dependent Hartree-Fock (TDHF), so that time dependent mean fields do not
prevent us from going on. We start from the TDHF equation with a specific guess for the time dependent mean field, 
\begin{equation}
\left( -i \gamma_5 \partial_x + m \gamma^0 e^{2ibt\gamma_5} \right) \psi = i \partial_t \psi.
\label{A29}
\end{equation}
This temporal chiral spiral implies the scalar and pseudoscalar potentials 
\begin{equation}
S(t) = m \cos (2bt), \quad P(t)= m \sin(2bt).
\label{A30}
\end{equation}
The mean field is constant in space, but rotates uniformly in time around the chiral circle with angular frequency $\Omega=2b$. 
This seems to fit nicely into the concept of ``spontaneous symmetry probing" of Refs.~\cite{L15d,L15e}.
Following the steps above, we
have to verify self-consistency, determine the free parameters $m,b,$ and evaluate the axial density and energy density.
But first of all we need to solve the TDHF equation (\ref{A29}). The time dependent, unitary transformation into a rotating frame,
\begin{equation}
\psi = e^{-ibt\gamma_5} \tilde{\psi},
\label{A31}
\end{equation} 
eliminates the explicit time dependence from the TDHF equation,
\begin{equation}
\left( - i \gamma_5 \partial_x - b \gamma_5 + m \gamma^0 \right) \tilde{\psi} = i \partial_t \tilde{\psi}.
\label{A32}
\end{equation}
We can then proceed as in the HF case with the 
stationary ansatz
\begin{equation}
\tilde{\psi} = e^{-i\epsilon t}\phi ,
\label{A33}
\end{equation}
reducing Eq.~(\ref{A32}) to a time independent Dirac equation for $\phi$,
\begin{equation}
\left( - i \gamma_5 \partial_x - b \gamma_5 + m \gamma^0 \right) \phi = \epsilon  \phi.
\label{A34}
\end{equation}
Its solutions can again be inferred from those of the free, massive Dirac equation, but now with shifted momenta,
\begin{eqnarray}
\phi^{(+)} & = &  e^{ikx} u_{k-b}, \quad \epsilon= e_{k-b},
\nonumber \\
\phi^{(-)} & = &  e^{ikx} v_{k-b}, \quad \epsilon= - e_{k-b}. 
\label{A35}
\end{eqnarray}
Hence the solutions of the original TDHF equation (\ref{A29}) are  
\begin{eqnarray}
\psi_k^{(+)} & = &  e^{-ibt\gamma_5} e^{-i e_{k-b}t} e^{ikx}  u_{k-b} ,
\nonumber \\
\psi_k^{(-)} & = &  e^{-ibt\gamma_5} e^{i e_{k-b}t} e^{ikx}  v_{k-b} .
\label{A36}
\end{eqnarray}
We occupy all $\psi_k^{(-)}$ states in the interval $k\in [-\Lambda/2, \Lambda/2]$, just like in the vacuum.
In order to check self-consistency and determine the free parameters, we note that the 
scalar and pseudoscalar densities for the level $\psi_k^{(-)}$ read
\begin{eqnarray}
\bar{\psi}_k^{(-)} \psi_k^{(-)} & = & - \cos (2bt) \frac{m}{e_{k-b}},
\nonumber \\
\bar{\psi}^{(-)} i \gamma_5 \psi_k^{(-)} & = &- \sin(2bt) \frac{m}{e_{k-b}}.
\label{A37}
\end{eqnarray}
Evaluation of the mean fields,
\begin{eqnarray}
S & = &  - Ng^2 \int_{-\Lambda/2}^{\Lambda/2} \frac{dk}{2\pi} \bar{\psi}_k^{(-)} \psi_k^{(-)}
 =   \frac{Ng^2}{2\pi} \ln \left( \frac{\Lambda^2}{m^2}\right) m \cos (2bt),
\nonumber \\
P & = &  - Ng^2 \int_{-\Lambda/2}^{\Lambda/2} \frac{dk}{2\pi} \bar{\psi}_k^{(-)} i \gamma_5 \psi_k^{(-)}
 =   \frac{Ng^2}{2\pi} \ln \left( \frac{\Lambda^2}{m^2} \right) m \sin(2bt).
\label{A38}
\end{eqnarray}
Once again this agrees with the input (\ref{A30}) owing to the vacuum gap equation, provided we set $m=m_0$.
The axial charge density is readily computed,
\begin{equation}
\frac{\rho_5}{N}  =  \int_{-\Lambda/2}^{\Lambda/2}   \frac{dk}{2\pi} \psi_k^{(-)\dagger} \gamma_5 \psi_k^{(-)} =
- \int_{-\Lambda/2}^{\Lambda/2} \frac{dk}{2\pi}  \frac{(k-b)}{e_{k-b}}=  \frac{b}{\pi}.
\label{A39}
\end{equation}
This provides us with the relation between the rotational frequency of the mean field and the axial density. Finally we
compute the energy density,
\begin{equation}
\frac{\cal E}{N}  =  \int_{-\Lambda/2}^{\Lambda/2} \frac{dk}{2\pi} \psi^{(-)\dagger}_k h \psi_k^{(-)} + \frac{m_0^2}{2Ng^2} ,
\label{A40}
\end{equation}
with the time dependent HF single particle Hamiltonian
\begin{equation}
h = \left( -i \gamma_5 \partial_x + m_0 \gamma^0 e^{2ibt\gamma_5} \right) .
\label{A41}
\end{equation}
We find
\begin{eqnarray}
\psi_k^{(-)\dagger} h \psi_k^{(-)} & = & - \frac{k^2+m_0^2-kb}{e_{k-b}},
\nonumber \\
\frac{\cal E}{N} & = & \frac{{\cal E}_0}{N}+ \frac{b^2}{2\pi}.
\label{A42}
\end{eqnarray}
\begin{figure}[h]
\begin{center}
\epsfig{file=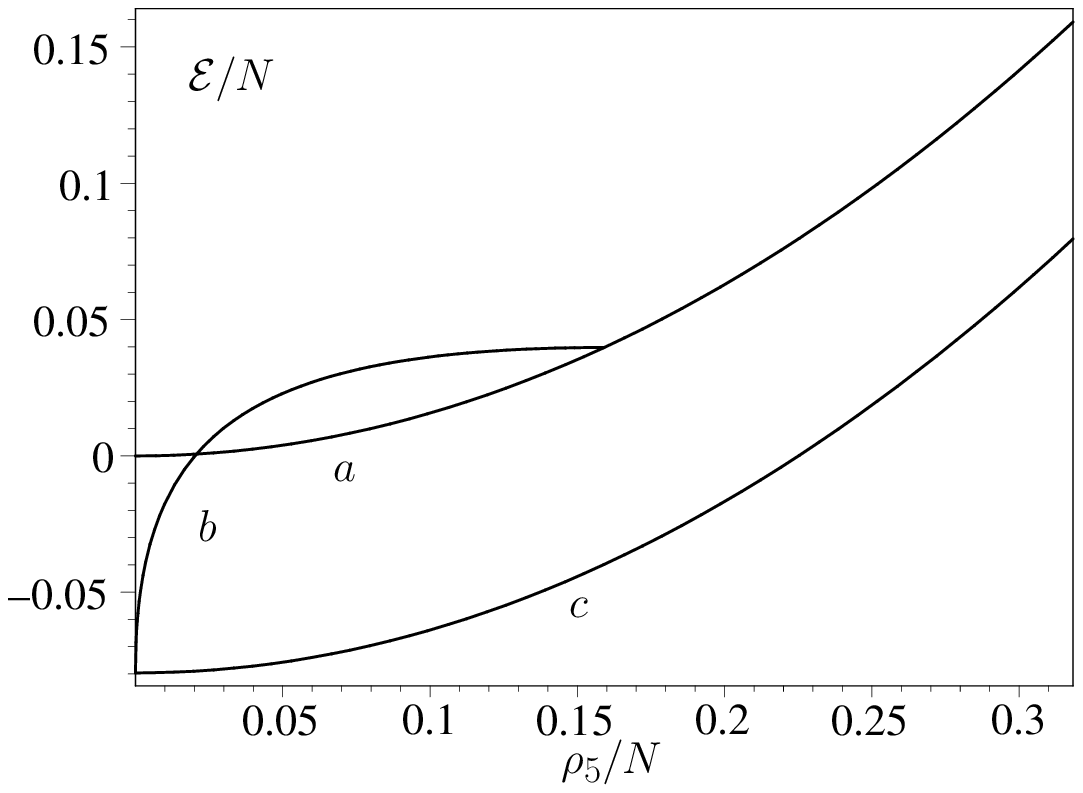,width=8cm,angle=0}
\caption{Energy density vs. $\rho_5$ for NJL$_2$ model at finite axial density, in units where $m_0=1$ . Curves $a$ and $b$ are the same as in Fig.~\ref{Fig3}, but plotted against $\rho_5$.
Curve $c$ is the result of the temporal chiral spiral, Eq.~(\ref{A42}).}
\label{Fig4}
\end{center}
\end{figure}
Hence the axial density and energy density relative to the vacuum are $b/\pi$ and $b^2/2\pi$, strongly reminiscent of 
the finite density results. If we want to compare this TDHF solution with the one assuming a time independent mean field,
we must take into account the different relation between the axial charge density and the parameters $k_f$ and $b$,
respectively. We therefore plot both energy densities as a function of $\rho_5/N$ in Fig.~{\ref{Fig4}. This plot shows that the time dependent
mean field solution is energetically favored at all densities. In the same way as the chiral spiral breaks translational
invariance in the finite density case, we conclude that time translation symmetry is spontaneously broken if one
considers the ground state in a sector with finite axial charge density.

So far, we have discussed the two cases of the NJL$_2$ model with pure fermion density or pure current density. More generally,
one could prescribe two non-vanishing parameters $(\rho,\rho_5)$ and ask for the ground state of this system. Since this situation
in turn is related by a Lorentz boost to either of the two special cases, we can predict the generalized chiral spiral without further
calculation. The mean field
\begin{equation}
m_0 \gamma^0 e^{-2i(ax - bt)\gamma_5}
\label{A43}
\end{equation}
will induce the densities $\rho=a/\pi, \rho_5=b/\pi$. If $|\rho|>|\rho_5|$, it describes a spatial chiral spiral moving with rapidity $\xi = {\rm artanh}\, \rho_5/\rho$. If $|\rho_5|>|\rho|$, it corresponds
to a temporal chiral spiral moving with rapidity $\xi = {\rm artanh}\, \rho/\rho_5$. It is natural to refer to these two cases as ``spacelike" and ``timelike" chiral spirals.
Fig.~\ref{Fig5} illustrates which points in the ($\rho,\rho_5$)-plane are related by Lorentz boosts and where spacelike and timelike chiral spirals
can be found.
The energy and momentum densities of such ``space-time crystals" can be evaluated along the same lines as above with the covariant result
\begin{equation}
\frac{\cal E}{N} = \frac{a^2+b^2}{2\pi}, \quad \frac{\cal P}{N} = \frac{ab}{\pi},
\label{A44}
\end{equation}
where we have  used the cutoff $\Lambda'$ defined below Eq.~(\ref{A20}) and subtracted the vacuum energy density.
\begin{figure}[h]
\begin{center}
\epsfig{file=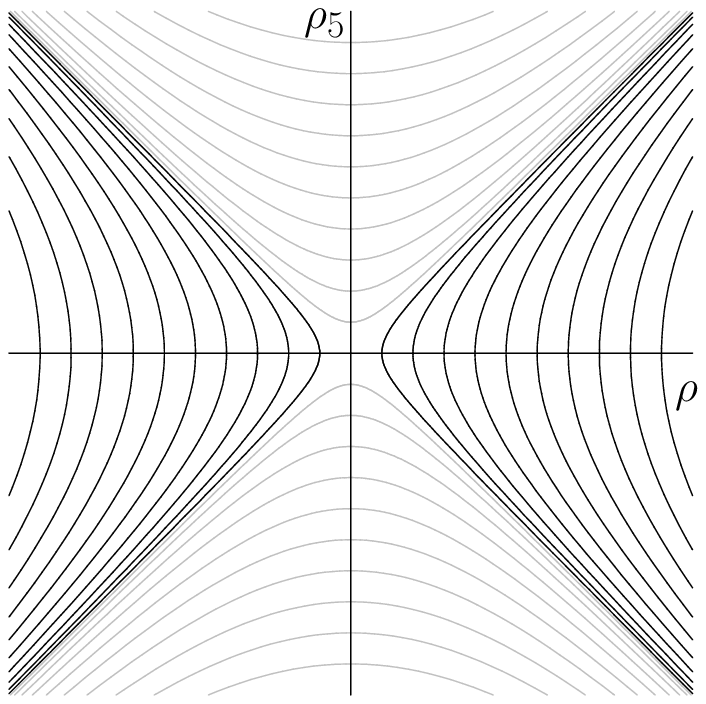,width=8cm,angle=0}
\caption{Foliation of the ($\rho,\rho_5$)-plane with curves connecting points related by Lorentz boosts. Black lines: spacelike chiral spirals, grey lines: timelike chiral spirals.}
\label{Fig5}
\end{center}
\end{figure}

An interesting question is what happens at finite temperature. In the conventional approach, one would answer this question
with the help of the grand canonical potential $H-\mu Q - \mu_5 Q_5$ with distinct vector and axial vector chemical potentials $\mu, \mu_5$.
Such studies have indeed been performed in various field theoretic models \cite{L16,L17,L18,L19,L20}. From our results at $T=0$ we would expect
that time dependent order parameters should also arise at finite temperature. We leave the question about the phase structure of the NJL$_2$ model
as a function of $T, \mu, \mu_5$ for the future.

Finally, we expect that the zero temperature results for the NJL$_2$ model carry over to the 't~Hooft model \cite{L21} (SU($N_c$) QCD
with massless quarks in 1+1 dimensions in the large $N_c$ limit) as well. For the case of finite density this has been shown
in \cite{L12} and verified independently in \cite{L22}. As a matter of fact, it should hold in all fermionic models where the interaction
has a local chiral symmetry, since all non-trivial effects then originate from the kinetic term of the Lagrangian \cite{L15}. 
The main reason why we have chosen the NJL$_2$ model is the fact that the vacuum and the homogeneous phases can be determined
in closed analytical form here, whereas numerical calculations are necessary in the 't~Hooft model.

The author would like to thank Gerald Dunne for helpful comments and suggestions.

\end{document}